**Title:** The Complex Estimand of Clone-Censor-Weighting When Studying Treatment Initiation Windows

**Authors:** Michael Webster-Clark,[1,2] Yi Li,[1] Sophie Dell'Aniello,[3] and Robert W. Platt[1,3]

**Affiliations:**

1. Department of Epidemiology, Biostatistics, and Occupational Health, McGill University, Montreal, QC, CA
2. Department of Epidemiology, University of North Carolina at Chapel Hill, Chapel Hill, NC, USA
3. Lady Davis Institute, Jewish General Hospital, Montreal, QC, CA.

**ABSTRACT:**

Clone-censor-weighting (CCW) is an analytic method for studying treatment regimens that are indistinguishable from one another at baseline without relying on landmark dates or creating immortal person time. One particularly interesting CCW application is estimating outcomes when starting treatment within specific time windows in observational data (e.g., starting a treatment within 30 days of hospitalization). In such cases, CCW estimates something fairly complex. We show how using CCW to study a regimen such as "start treatment prior to day 30" estimates the potential outcome of a hypothetical intervention where A) prior to day 30, everyone follows the treatment start distribution of the study population and B) everyone who has not initiated by day 30 initiates on day 30. As a result, the distribution of treatment initiation timings provides essential context for the results of CCW studies. We also show that if the exposure effect varies over time, ignoring exposure history when estimating inverse probability of censoring weights (IPCW) estimates the risk under an impossible intervention and can create selection bias. Finally, we examine some simplifying assumptions that can make this complex treatment effect more interpretable and allow everyone to contribute to IPCW.

Researchers have been attempting to mimic randomized trials in nonexperimental data for decades. One persistent challenge has been correctly allocating the time experienced by individuals while different treatment regimens cannot be distinguished from one another. Randomized trials can assign patients to different durations of treatment (e.g., randomize patients receiving chemotherapy for colon cancer to receive up to six vs or up to twelve infusions). In observational data, on the other hand, it can be impossible to determine the treatment regimens compatible with a specific patient's data until well after baseline. Traditional means to analyze these data generally change the estimated quantity vs a hypothetical trial,(1) require strong assumptions to estimate that quantity,(2) or fail to align the dates when eligibility is assessed, treatment starts, and follow-up begins.(3, 4)

One analytic solution available in nonexperimental data involves **cloning** participants into various treatment regimens, **censoring** clones when they are no longer compatible with that treatment regimen, and **weighting** uncensored clones based on the inverse of their probability of remaining uncensored (IPCW) (hereafter referred to as CCW for clone-censor-weighting).(5) Initially proposed to study dynamic treatment regimens, CCW has become closely tied to the target trial framework as a method for studying treatment regimens indistinguishable from one another at baseline without randomization or outcome modeling with the G-formula.(6, 7) In addition to the regimen types above, CCW can compare outcomes when individuals receive the same treatment at different times (e.g., surveillance colonoscopy every 3 vs. 7 years).(8) CCW can also study the effect of initiating treatment within a specific window (e.g., statin initiation within 6 months vs no treatment(9) or starting hormone therapy at 3 months vs at disease progression).(10)

Unfortunately, the quantity CCW estimates when used to study regimens like "start a treatment within 30 days" can be difficult to articulate. Without major simplifying assumptions, it does not correspond to everyone starting at day 0, or day 30, or randomly from day 0 to 30, or randomly depending on covariates from day 0 to 30. Using simple data sets of a few individuals and a straightforward simulation, we demonstrate that using CCW to study a regimen such as "start treatment prior to day 30" estimates potential outcomes under a hypothetical intervention where A) prior to day 30, everyone follows the treatment start distribution of the study population and B) everyone who has not initiated by day 30 is forced to initiate on day 30. We also show that CCW that omit cumulative exposure history treat everyone who is forced to initiate on day 30 as if they have the cumulative exposure history of those who initiated prior to day 30 (resulting in an impossible intervention or selection bias, depending on your perspective) and that this problem is avoided if the only individuals contributing to the IPCWs are those who initiate on day 30. Finally, we show some simplifying assumptions that make it easier to interpret this complicated quantity.

**METHODS**

**Key context**

*Primary goal*-We aimed to articulate and describe the hypothetical intervention that would generate the quantity estimated when applying CCW methods to a specific population and studying a treatment regimen such as "add inhaled corticosteroids (ICS) to long-acting beta-agonists (LABA) within the 30 days following discharge from the hospital with a primary diagnosis of chronic obstructive pulmonary disease (COPD) exacerbation" in situations without any simplifying assumptions (e.g., the absence of a treatment effect until 30 days or otherwise treating day 0 and day 30 as completely interchangeable). We focused on the expectation of a binary potential outcome (e.g., $E[Y^{a=1}]$) under a single hypothetical intervention, rather than a treatment effect involving contrasting two potential outcomes under corresponding interventions on a specific scale (e.g., $E[Y^{a=1}] - E[Y^{a=0}]$).

*An overview of using CCW to study different timings of treatment initiation*-Implementing CCW can be daunting, but it is not as complex as it might initially seem.(3, 8) Take the above example estimating potential outcomes in those who add ICS to LABA after hospitalization for a COPD exacerbation.

**First,** specify the index event, eligibility criteria, and the treatment regimen (which, identify the censoring criteria). In the example, the index event would be "discharge from the hospital after COPD exacerbation," the eligibility criteria would be "current use of LABA and no past use of ICS," and the treatment regimen would be "initiate ICS within 30 days of the index event." This regimen has a single censoring criteria: not initiating ICS within 30 days of the index event, meaning that no one will be censored until day 30.

**Second,** identify everyone who A) experienced the index event and B) met eligibility criteria. Create a copy, or **clone**, of these individuals assigned to each treatment regimen of interest. For the COPD study, this would mean finding all individuals that are 1) discharged from the hospital with a COPD exacerbation, 2) are currently taking LABA, and 3) have no past use of ICS. Each person meeting the criteria is cloned, with the clone assigned to the "initiate ICS within 30 days of the index event" treatment regimen. Note that we are examining only one treatment regimen and one potential outcome and could use the original observations; cloning is mentioned here for those unfamiliar with the design.

**Third,** apply censoring criteria to the clones within the corresponding treatment regimen. In the example, that would mean censoring any clones from the data set who did not initiate treatment with ICS on or prior to day 30 on day 30.

**Fourth,** to handle potential selection bias created by this censoring process, create time-varying weights for each clone's person-time based on the covariate-conditional inverse probability of remaining uncensored (IPCW).(11, 12) Additional censoring processes (e.g., loss of insurance coverage or changing healthcare providers) must be handled separately. As a note, if the goal is to estimate incidence rates and there is an effect of treatment, weights must be used **even if** no covariates predict censoring. In our case, this would mean identifying the clones who survived to day 30, estimating the probability of remaining uncensored at day 30, and then assigning the clones who remained uncensored IPCW.

**Figure 1** is a visual representation of steps two, three, and four. We create new copies of everyone eligible for the analysis (in the figure, individuals 2, 3, and 4), starting with the index date; we censor the observations when they deviate from our "start ICS by 30 days" regimen (meaning clone 3 and 4 get censored at day 30); and then we create IPCW, where each clone receives a weight of 1 for the first 30 days and then clone 2 receives a weight of 3 starting at day 30 because that is their inverse probability of remaining uncensored.

Finally, all that is left is to analyze the data to obtain the outcome measure of interest (e.g., risk or incidence rate) using a method that appropriately accounts for the weights we created.

**A simple demonstration of the CCW intervention**

With this process clearly understood, let us apply CCW to three different small data sets shown in **Figure 2**.

*Simple data set 1 (**Figure 2A**)*-This data set includes only two individuals: Adrianna and Betty. Adrianna never starts ICS, while Betty starts ICS exactly 30 days following her index hospitalization. We clone both individuals and assign them to the "treat by 30 days" regimen. Adrianna's clone is censored at day 30, while Betty's clone is never censored. This means that after day 30 Betty's person-time receives a weight of 2 so that their outcomes "stand in" for Adrianna's. Note that if we had more people, we might instead only upweight those individuals with covariates similar to Adrianna at day 30. Since we keep Adrianna in our analytic data set prior to day 30, it's as though they initiated ICS on day 30. Using CCW to study a 30-day initiation window with this data set corresponds to an intervention where Adrianna and Betty both start treatment on day 30.

*Simple data set 2 (**Figure 2B**)*-This data set still includes Adrianna, but includes a new individual (Chen) rather than Betty. Instead of starting treatment on day 30, like Betty did, Chen started ICS immediately following hospital discharge (meaning they started on day 0). We create our two clones and once again assign them to the "start ICS by 30 days" regimen. Chen's clone is never censored and Adrianna's clone is, once again, censored on day 30. Just like Betty, Chen receives a weight of 2 to "stand in" for Adrianna after day 30. Since we keep Adrianna in the analytic data set prior to day 30, it is still as if Adrianna initiated ICS on day 30. This time, using CCW to study a 30-day initiation window corresponds to an intervention where Chen starts treatment at day 0 and Adrianna starts treatment at day 30. More than that, because we have no choice but to use Chen to "stand in" for Adrianna following day 30, it's as though Adrianna started treatment on day 30 **with the previous exposure profile of Chen** (i.e., as if they were on treatment since day 0).

This is **impossible**; we cannot start Adrianna on ICS on day 30 as if she were exposed to ICS since day 0. This can also be viewed as a form of selection bias, with exposure history predicting the probability of remaining uncensored and risk of the outcome.(13-15) In general, if sample data set 2 represents all our data, we cannot emulate a real trial involving the "start by 30 days" treatment regimen without strong assumptions about the treatment effect (to be discussed below).

*Simple data set 3 (**Figure 2C**)*-The final simple data set includes Adrianna, Betty, and Chen, meaning it includes one individual who started ICS immediately after discharge (Chen), one individual who started 30 days after discharge (Betty), and one person who never started at all (Adrianna). Once again, Betty and Chen are never censored and Adrianna is censored at day 30. Rather than receiving a weight of 2 after day 30, however, Betty and Chen now each receive a weight of 1.5 (the two of them "stand in" for Adrianna as a pair). Our hypothetical intervention again starts Chen on day 0 and Adrianna and Betty on day 30, but now treats Adrianna's exposure history after day 30 as if it were an equal mixture of Betty's (i.e., no previous exposure) and Chen's (i.e., exposed since day 0). Once again, this is an **impossible intervention**, as our theoretical trial would have to start Adrianna on day 30 but somehow intervene to set her exposure history to be equally no exposure (from Betty) and 30 prior days of exposure (from Chen).

*Avoiding impossible interventions and selection bias*-Fortunately, there is a way to ensure IPCW do not estimate the effect of an impossible intervention when the data include a mix of individuals who started at the end of the exposure window and individuals who started earlier- **estimate the probability of remaining uncensored at day 30 only among those who either initiated at day 30** (or, if there are not enough people on exactly day 30, those close enough in time to day 30 such that they have equivalent exposure history) **or did not.** Those who initiated at day 30 then receive an IPCW based on this probability and everyone who initiated prior to day 30 receives an IPCW of 1 (or, if there's other censoring criteria, the weight they had previously)**.** In data set 3, for example, we can give Chen's follow-up after day 30 a weight of 1 and Betty's follow-up after day 30 a weight of 2 so that the only one "standing in" for Adrianna is Betty. We are now estimating the effect of starting Chen on day 0 and Adrianna and Betty on day 30, an intervention that could be implemented in a real trial.

Notably, that the other intervention is impossible does not mean it is always uninformative. Under specific assumptions, the outcome under the impossible intervention equals the outcome under the more reasonable one; it is only problematic if exposure history and timing is an outcome predictor independent of current exposure. If there is no treatment effect or the effect of exposure is instantaneous and constant (i.e., similar to diuretics' effect on blood pressure or heparin's effect on preventing blood clots), exposure history will not predict the outcome or create selection bias and the treatment effect estimated will be the same whether IPCW in data set 3 include both Betty and Chen or only Betty.

*A general rule for the intervention*-With these examples in mind, using CCW to study a "start by 30 days" regimen corresponds to a two-stage intervention:

1. Until day 30, allow the population to initiate treatment under the "natural course" of the original population **observed from day 0 to 30.**
2. Force individuals in the population who have yet to initiate treatment at day 30 to start immediately.

Thus, the treatment distribution underlying the hypothetical intervention varies depending on how initiation timing is distributed in the study population.

Notably, when studying a treatment regimen such as initiating treatment from 30 to 90 days, the hypothetical intervention has three steps:

1. Until day 30, prevent anyone in the population from initiating.
2. From day 30 to 90, allow the population to initiate treatment under the "natural course" of the original population **observed from day 30 to 90**.
3. Force any individuals in the population who have yet to initiate treatment at day 90 to start immediately.

**Simulation description**

To verify the results of the simple examples in more complex situations and test the extent to which assumptions could simplify the intervention discussed above, we created several different simulated data sets with a time-varying treatment X that was continued once started, an outcome Y, and another covariate C that was associated with the outcome. Outcomes occurred over three time units, each representing a month in our example with inhaled corticosteroids (i.e., period 0-1 = day 0-30, 1-2 = day 30-60, 2-3 = day 60-90). The only source of censoring is individuals failing to initiate treatment by day 30.

*Base simulation*-Our base simulation mimicked the situation with ICS, meaning that X had an effect on Y that occurred immediately but increased over time, with an increase after the first time unit. The mathematical model for the outcome was:

$$P(Y) = (intercept) + 0.1 * C + 0.1 * X + 0.04 * CumulativeX + 0.05 * previous\, X.$$

*Additional scenarios*-We simulated five other scenarios exploring how specific types of exposure effects impacted the similarity of different CCW estimates to one another and to the various interventions. These included:

A) A scenario with no exposure effect.

B) A scenario where exposure had a purely instantaneous effect.

C) A scenario where anyone treated at time 0 or 1 experienced an effect starting at time 1 (regardless of whether an individual initiated at time 0 or time 1).

D) A scenario where exposure had a delayed effect that begins 1 unit after initiating treatment.

E) A scenario where the effect of exposure was purely cumulative over time.

*CCW analyses:* The hypothetical regimen examined was "treat by time 1", meaning individuals were censored from the regimen if they did not initiate at time 0 or time 1. We performed two separate CCW analyses. In one analysis ("limited" CCW), only those who initiated at time 1 were included in IPCW, with all the time 0 initiators receiving weights of 1. In the other ("all initiator" CCW), we allowed those who initiated at time 0 to contribute to IPCW estimated at time 1. We estimated the time 3 (i.e., 90 day) risks using both approaches.

*Hypothetical interventions:* We compared CCW time 3 risks to the true risks obtained via Monte Carlo simulation(16) with 10 repeated samples of the original 5,000,000 individual population using the true treatment and outcome models under four different hypothetical interventions. These interventions included:

1. Start everyone on X at time 0.

2. Start everyone on X at time 1.

3. Allow the population to follow its natural exposure distribution until time 1 and then start everyone who has yet to start by time 1 to start at time 1 ("feasible intervention").

4. Allow the population to follow its natural exposure distribution until time 1 and then start everyone who has yet to initiate by time 1 at time 1, assigning those who are forced to start in this way a previous exposure history based on the distribution among those who naturally started by time 1 ("impossible intervention").

In every scenario, the probability of starting X at time 0 was higher in those with C=1 and the probability of Y was higher in those with C=1 (making adjustment for C necessary to estimate an unbiased risk under any intervention). Each scenario included 5,000,000 individuals to minimize the influence of random variation. **Figure 3** summarizes the simulation structure and interventions we examined. Full parameters for the outcome model in each scenario are listed in **Table 1**.

**RESULTS**

**Table 2** shows each scenario's time 3 risk in the cohort after cloning, censoring, and using either "limited" CCW that only upweight those who started at time 1 or "all initiator" CCW that include those who started at time 0 and time 1. The panels also include the time 3 risk when implementing the four interventions.

In the base case, all four interventions yield different risks, but the "limited" (56.6%) and "all initiator" (risk=59.1%) CCW approaches matched the "feasible" (56.6%) and "impossible" (59.1%) interventions, respectively. When there is no effect of the treatment (Scenario A) or when the exposure effect takes effect at time 1 regardless of whether it is received at time 0 or time 1 (Scenario C), all interventions and CCW approaches result in the same risk estimate (26.4% or 41.8%, respectively). When the exposure effect is only instantaneous in Scenario B, on the other hand, the two CCW approaches and the "impossible" and "feasible" interventions still yield the same risk (~43.1%), but that risk differs from the interventions that start everyone at time 0 (48.2%) or time 1 (41.8%). If the treatment effect is cumulative as in scenario D or takes effect after one time unit as it does in scenario E (i.e., when the exposure effect is time-varying), however, all four interventions result in different risks of the outcome, with the "all initiator" CCW analysis again corresponding to the risk of the "impossible" intervention.

**Table 3** summarizes how different situations impact whether A) one can include those who initiate before the end of the period in IPCW and B) whether the distribution of treatment during the period can be safely ignored.

**DISCUSSION**

Cloning, censoring, and weighting individuals can estimate potential outcomes under regimens that involve initiating treatment in a specific window. That said, the hypothetical intervention it corresponds to (having the population naturally initiate treatment prior to that date and then forcing everyone remaining to initiate on that date) is quite complicated and population specific. While it does correspond to a hypothetical target trial, that randomized trial may not align well with a relevant research question. In particular, the intervention can correspond to quite different exposure patterns in different populations during the treatment initiation window, with **Figure 4** providing examples of how cumulative initiation patterns under this hypothetical intervention can change in different populations. Perhaps most critically, if IPCW contributors are selected inappropriately, the intervention can diverge from any realistic intervention by retroactively assigning those who initiate at the end of the treatment window the exposure history of those who initiated prior. Still, some assumptions like there being no effect or no effect until the end of the time window can render the outcome estimated by CCW more interpretable and not limited to the specific initiation pattern present in the study population.

Clarifying the quantity estimated by CCW in this context has major advantages. First, it establishes that researchers using the design to study different treatment windows should provide information on the distribution of treatment timings within each treatment window to provide context to their results. If everyone in the study population initiates on day 89 or 91, differences between "start by day 90" and "start from day 90-180" regimens are very small, while in another population in which everyone starts on either day 0 or 180 differences could be enormous. It also reinforces the importance of avoiding comparisons like "start by day 7" to "do not start by day 7", as the latter includes everything from starting on day 8 to never starting. Finally, it stresses the importance of the study's time axes; while a 7-day window may create too much variation for a study of the 14-day effectiveness of antibiotics (rapid effect onset and short treatment duration), a 30-day window may be completely interchangeable when studying five-year effectiveness of statins (long onset of effect and long duration of treatment).

Second, understanding the exact quantity estimated by CCW is essential for research using simulations to contrast the performance of CCW with alternatives for estimating the effect of treatment regimens that are indistinguishable at baseline (e.g., the G-formula).(17, 18) In any but the simplest simulation studies, "true" potential outcomes (and any associated treatment effects) come from Monte Carlo simulation of the relevant intervention using the true treatment, covariate, and outcome models. This "true" outcome can be used to calculate bias, mean squared error, and confidence interval coverage to compare with other analytic methods.(18) Understanding the relevant assumptions also helps identify when researchers can derive the truth from simpler Monte Carlo simulations that do not incorporate the specific timing of treatment initiation or cumulative exposure.

We also identified when CCW inadvertently estimates an impossible treatment effect involving treating people who initiate at the end of the window as if they had previous exposure to treatment. This selection bias can be avoided if IPCW are created in initiators from the end of the interval, with those who initiated previously still contributing to the study but keeping their previous weights (similar, but not identical to, to fixing weights at 1)(19). While this approach may not always be required, researchers should consider it when the treatment effect under study varies over time due to a lag in treatment effect, a cumulative exposure effect, a delayed exposure effect, or other mechanisms. The appropriate definition of "end of the window," the precision lost by excluding others from IPCW, and whether this matters at all depends on a study-specific factors ranging from the distribution of times when people begin to be exposed to the exposure itself to the outcome of interest.

This work is not comprehensive. We focused on a single application of CCW (the study of initiating treatment within a specific window from an index event) and on the hypothetical intervention underlying the outcome it estimates. We also used a very simple simulation with one static confounder. While there are some situations that closely parallel this one (such as stopping a sustained treatment within a specific time window),(20) they were not the focus of this research. Future work exploring elements unique to comparing potential outcomes under hypothetical interventions (e.g., confounding, effect measure modification, and time-varying confounding) is essential to lend additional interpretability to risk differences, risk ratios, incidence rate ratios, and hazard ratios obtained in studies using CCW. Methods for transporting or generalizing these effects to populations with different covariate distribution and different initiation timings also warrant investigation.

Finally, we hope that these results will not dissuade any researchers from conducting CCW analyses. Indeed, we think additional clarity surrounding CCW should only make people more willing to apply the method. While the hypothetical intervention may be complicated and may not correspond to a well-defined target trial without some assumptions, said assumptions may hold in many cases. Moreover, the population-specific nature of the resulting effect estimates is equally impactful for actual trials that randomize patients to start treatment within treatment windows.

*Conclusion*

Using CCW to estimate the consequences of starting a treatment in a specific window provides amazing opportunities for non-experimental epidemiologic research. If researchers do not properly account for the complexity of the quantity CCW estimates and do not share the timing of treatment during the treatment windows, however, they run the risk of misrepresenting their findings or estimating the effects of impossible interventions.

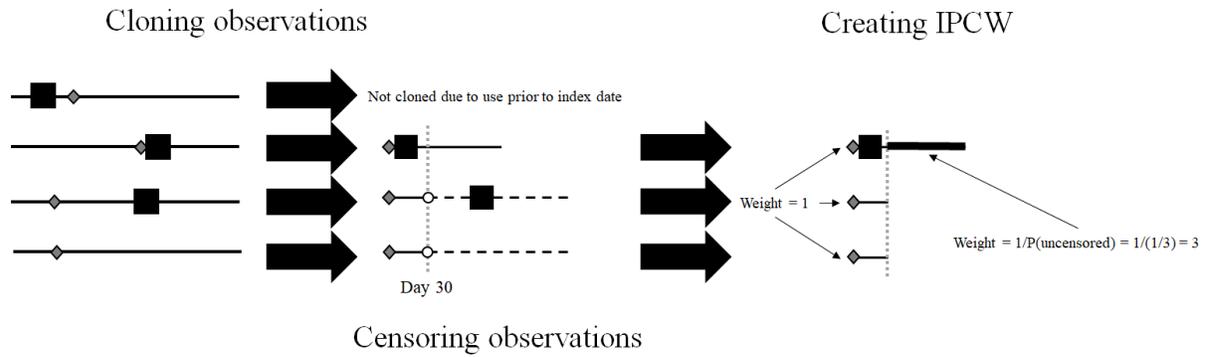

**Figure 1:** An overview of how to use cloning, censoring, and weighting to study an "initiate treatment by day 30 following an index event" regimen. Each line represents a potential observation, with gray diamonds corresponding to the index event, black boxes corresponding to the treatment that must be initiated within 30 days, and white circles corresponding to the date individuals are censored. The thickness of the line corresponds to the weight given to that section of person-time, and dashed lines represent time following a censoring event.

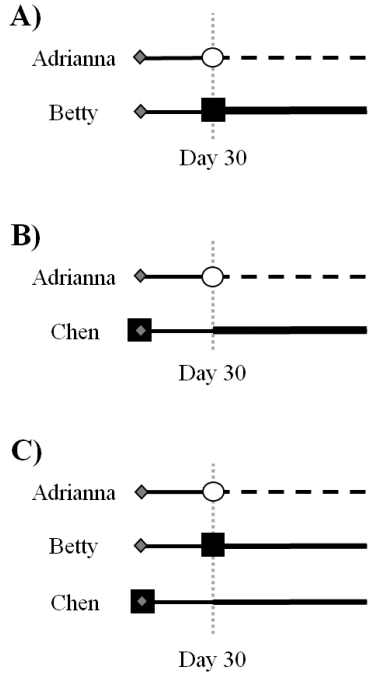

**Figure 2:** Three example data sets (panel A, B, and C) where we could implement clone-censor weighting to study a treatment regimen of initiating treatment (e.g., inhaled corticosteroids) by day 30 following an index event (e.g., COPD exacerbation) after meeting eligibility criteria (e.g., current use of LABA). The gray diamonds represent index events, the black squares the treatment of interest, and the white circles censoring events (in this case, not initiating by day 30). The thickness of the solid lines corresponds to the weight that time receives in the final analysis, with the dashed lines showing time following a censoring event.

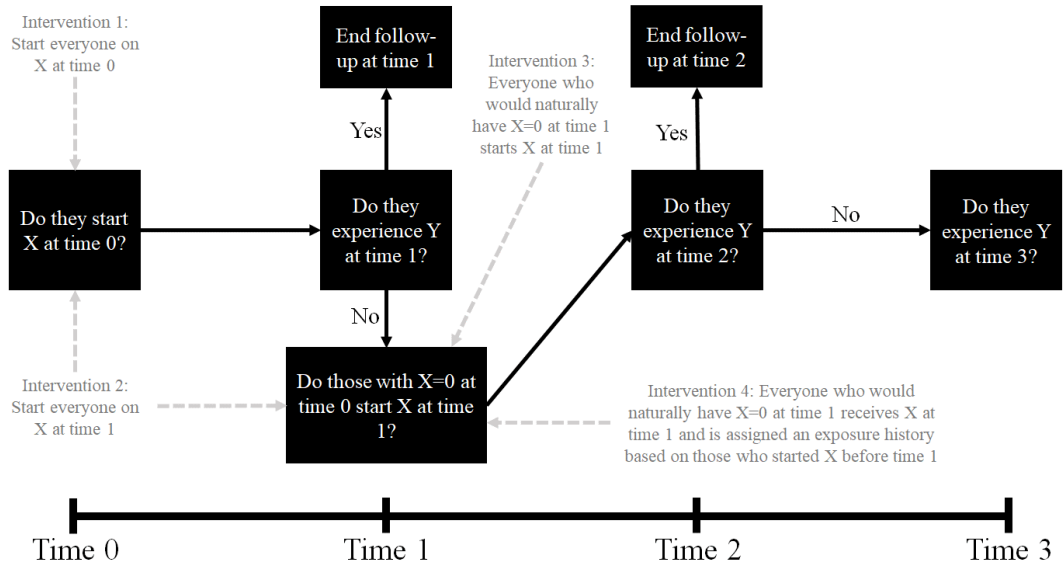

**Figure 3:** Simulation structure, with the black boxes representing different steps in the simulation process. Note that all individuals stay on X after starting it. The different hypothetical interventions are described using gray text with gray dashed arrows pointing to the steps and times an intervention would occur.

**Table 1:** Equations to calculate P(Y) for individuals in each scenario.

| Scenario | Equation for P(Y) |
|---|---|
| Base scenario | $P(Y) = 0.05 + 0.1 * C + 0.1 * X + 0.04 * (cumulative\ X) + 0.05 * (X\ in\ the\ last\ time\ period)$ |
| Scenario A: No exposure effect | $P(Y) = 0.05 + 0.1 * C$ |
| Scenario B: Exposure effect is instantaneous | $P(Y) = 0.05 + 0.1 * C + 0.1 * X$ |
| Scenario C: Exposure takes effect at time 1 | If t=0, $P(Y) = 0.05 + 0.1 * C$ <br> If t≥1, $(Y) = 0.05 + 0.1 * C + 0.1 * X$ |
| Scenario D: Exposure effect is delayed by 1 time unit | $P(Y) = 0.05 + 0.1 * C + 0.1 * (X\ in\ the\ last\ time\ period)$ |
| Scenario E: Exposure effect is cumulative | $P(Y) = 0.05 + 0.1 * C + 0.05 * (cumulative\ X)$ |

**Table 2:** Risk estimated at time 3 (i.e., 90 days) using different CCW methods along with the true risks obtained from Monte Carlo simulation under four hypothetical interventions.

| Scenario | Limited CCW | All initiator CCW | Hypothetical intervention 1: Start everyone at time 0 | Hypothetical intervention 2: Start everyone at time 1 | Hypothetical intervention 3: Feasible intervention[a] | Hypothetical intervention 4: Impossible intervention[b] |
|---|---|---|---|---|---|---|
| Base scenario | 56.6% | 59.1% | 67.4% | 53.6% | 56.6% | 59.1% |
| Scenario A: No exposure effect | 26.4% | 26.4% | 26.4% | 26.4% | 26.4% | 26.4% |
| Scenario B: Exposure effect is instantaneous | 43.2% | 43.2% | 48.2% | 41.8% | 43.1% | 43.2% |
| Scenario C: Exposure takes effect at time 1 | 41.8% | 41.8% | 41.8% | 41.8% | 41.8% | 41.8% |
| Scenario D: Exposure effect is delayed by 1 time unit | 36.1% | 38.6% | 41.8% | 34.6% | 36.1% | 38.6% |
| Scenario E: Exposure effect is cumulative | 40.3% | 42.8% | 48.4% | 38.2% | 40.4% | 42.8% |

CCW=clone censor weighting.

(a): An intervention allowing the population to follow its natural exposure distribution until time 1 and then starting everyone who has yet to start by time 1 to start at time 1.

(b) An intervention allowing the population to follow its natural exposure distribution until time 1 and then starting everyone who has yet to initiate by time 1 at time 1, assigning those who are forced to start in this way a previous exposure history based on the distribution among those who naturally started by time 1.

**Table 3:** Table describing the situations where estimates will be unbiased when including individual individuals who initiate prior to the end of the period and whether you can ignore the distribution of treatment during the initiation period when interpreting results.

| Situation | Can you include those who initiate prior to the end of the period in IPCW? | Can you ignore the distribution of treatment during the period when interpreting results? |
|---|---|---|
| No exposure effect | Yes | Yes |
| Exposure effect begins after the end of the period (regardless of when people start) | Yes | Yes |
| Exposure effect is instantaneous | Yes | No |
| Exposure effect is delayed (i.e., there is a lag period) | No | No |
| Exposure effect is cumulative or otherwise time-varying | No | No |

IPCW = inverse probability of censoring weights.

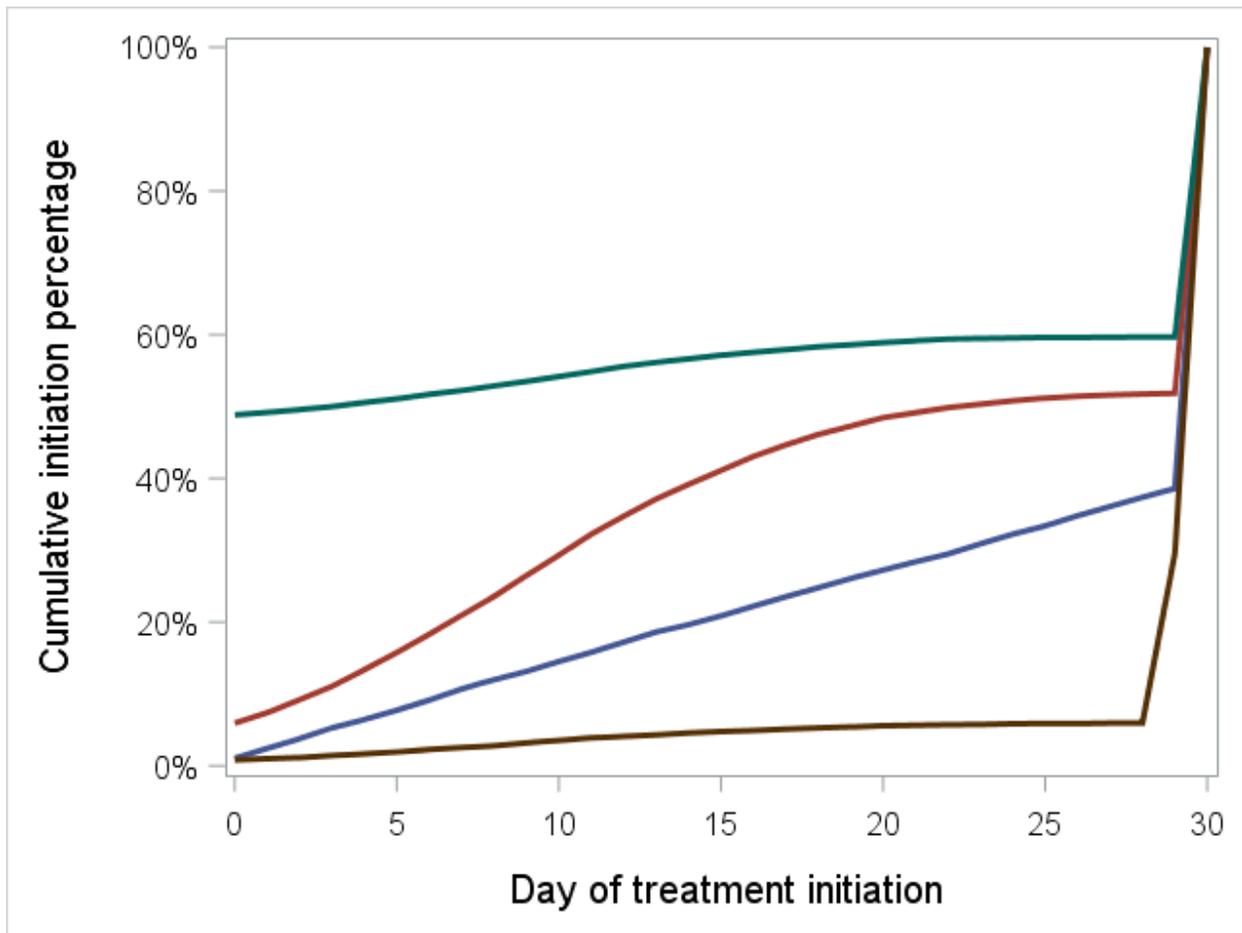

**Figure 4:** A plot of the cumulative proportion of the population initiating by each date from day 0 to day 30 in the intervention that corresponds to a CCW analysis of a "start by day 30" treatment regimen in various populations. Each line corresponds to a specific type of treatment distribution (brown = late start, green = early start, red = start times vary based on a normal distribution, blue = start times follow a uniform distribution). In every population, there is a sudden spike at day 30 when those who have yet to initiate start treatment under the hypothetical intervention.

# REFERENCES

1.	Rothman KJ, Suissa S. Exclusion of immortal person-time. Pharmacoepidemiology and drug safety. 2008;17(10):1036-.

2.	Suissa S. Immortal time bias in pharmacoepidemiology. American journal of epidemiology. 2008;167(4):492-9.

3.	Duchesneau ED, Jackson BE, Webster-Clark M, Lund JL, Reeder-Hayes KE, Nápoles AM, et al. The Timing, the Treatment, the Question: Comparison of Epidemiologic Approaches to Minimize Immortal Time Bias in Real-World Data Using a Surgical Oncology Example. Cancer epidemiology, biomarkers & prevention : a publication of the American Association for Cancer Research, cosponsored by the American Society of Preventive Oncology. 2022;31(11):2079-86.

4.	Hernán MA, Sauer BC, Hernández-Díaz S, Platt R, Shrier I. Specifying a target trial prevents immortal time bias and other self-inflicted injuries in observational analyses. Journal of clinical epidemiology. 2016;79:70-5.

5.	Cain LE, Robins JM, Lanoy E, Logan R, Costagliola D, Hernán MA. When to start treatment? A systematic approach to the comparison of dynamic regimes using observational data. The international journal of biostatistics. 2010;6(2).

6.	Hernán MA, Robins JM. Using Big Data to Emulate a Target Trial When a Randomized Trial Is Not Available. Am J Epidemiol. 2016;183(8):758-64.

7.	Hernán MA. Methods of Public Health Research - Strengthening Causal Inference from Observational Data. The New England journal of medicine. 2021;385(15):1345-8.

8.	Huitfeldt A, Kalager M, Robins JM, Hoff G, Hernán MA. Methods to estimate the comparative effectiveness of clinical strategies that administer the same intervention at different times. Current epidemiology reports. 2015;2:149-61.

9.	Emilsson L, García-Albéniz X, Logan RW, Caniglia EC, Kalager M, Hernán MA. Examining bias in studies of statin treatment and survival in patients with cancer. JAMA oncology. 2018;4(1):63-70.

10.	Garcia-Albeniz X, Chan J, Paciorek A, Logan R, Kenfield S, Cooperberg M, et al. Immediate versus deferred initiation of androgen deprivation therapy in prostate cancer patients with PSA-only relapse. An observational follow-up study. European Journal of Cancer. 2015;51(7):817-24.

11.	Robins JM, Hernan MA, Brumback B. Marginal structural models and causal inference in epidemiology. Epidemiology (Cambridge, Mass). 2000:550-60.

12.	Cain LE, Cole SR. Inverse probability-of-censoring weights for the correction of time-varying noncompliance in the effect of randomized highly active antiretroviral therapy on incident AIDS or death. Statistics in medicine. 2009;28(12):1725-38.

13.	Rothman KJ, Greenland S, Lash TL. Modern epidemiology: Wolters Kluwer Health/Lippincott Williams & Wilkins Philadelphia; 2008.

14.	Hernán MA. Invited Commentary: Selection Bias Without Colliders. Am J Epidemiol. 2017;185(11):1048-50.

15.	Daniel RM, Kenward MG, Cousens SN, De Stavola BL. Using causal diagrams to guide analysis in missing data problems. Statistical methods in medical research. 2012;21(3):243-56.

16.	Harrison RL. Introduction To Monte Carlo Simulation. AIP conference proceedings. 2010;1204:17-21.

17.	Mayeda ER, Tchetgen Tchetgen EJ, Power MC, Weuve J, Jacqmin-Gadda H, Marden JR, et al. A simulation platform for quantifying survival bias: an application to research on determinants of cognitive decline. American journal of epidemiology. 2016;184(5):378-87.

18.	Morris TP, White IR, Crowther MJ. Using simulation studies to evaluate statistical methods. Stat Med. 2019;38(11):2074-102.

19.	Fu EL, Evans M, Carrero J-J, Putter H, Clase CM, Caskey FJ, et al. Timing of dialysis initiation to reduce mortality and cardiovascular events in advanced chronic kidney disease: nationwide cohort study. BMJ (Clinical research ed). 2021;375:e066306.

20.	Larochelle MR, Lodi S, Yan S, Clothier BA, Goldsmith ES, Bohnert ASB. Comparative Effectiveness of Opioid Tapering or Abrupt Discontinuation vs No Dosage Change for Opioid Overdose or Suicide for Patients Receiving Stable Long-term Opioid Therapy. JAMA Network Open. 2022;5(8):e2226523-e.